\documentclass{tudelft-report}
\usepackage{float}
\usepackage{threeparttable}
\usepackage{booktabs}
\usepackage{setspace}
\usepackage{graphicx, subcaption}

\usepackage{wrapfig}

\usepackage{natbib} 
\usepackage{parskip}
\setlist{itemsep=-2pt} 

\usepackage{silence}
\begin{document}

\frontmatter

\title{Regional House Price Dynamics in Australia}
\subtitle{Insights into Lifestyle and Mining Dynamics in Australia's Housing Market through PCA}
\author{Will Sijp}

\subject{Neoval Report} 
\affiliation{Neoval} 

\coverimage{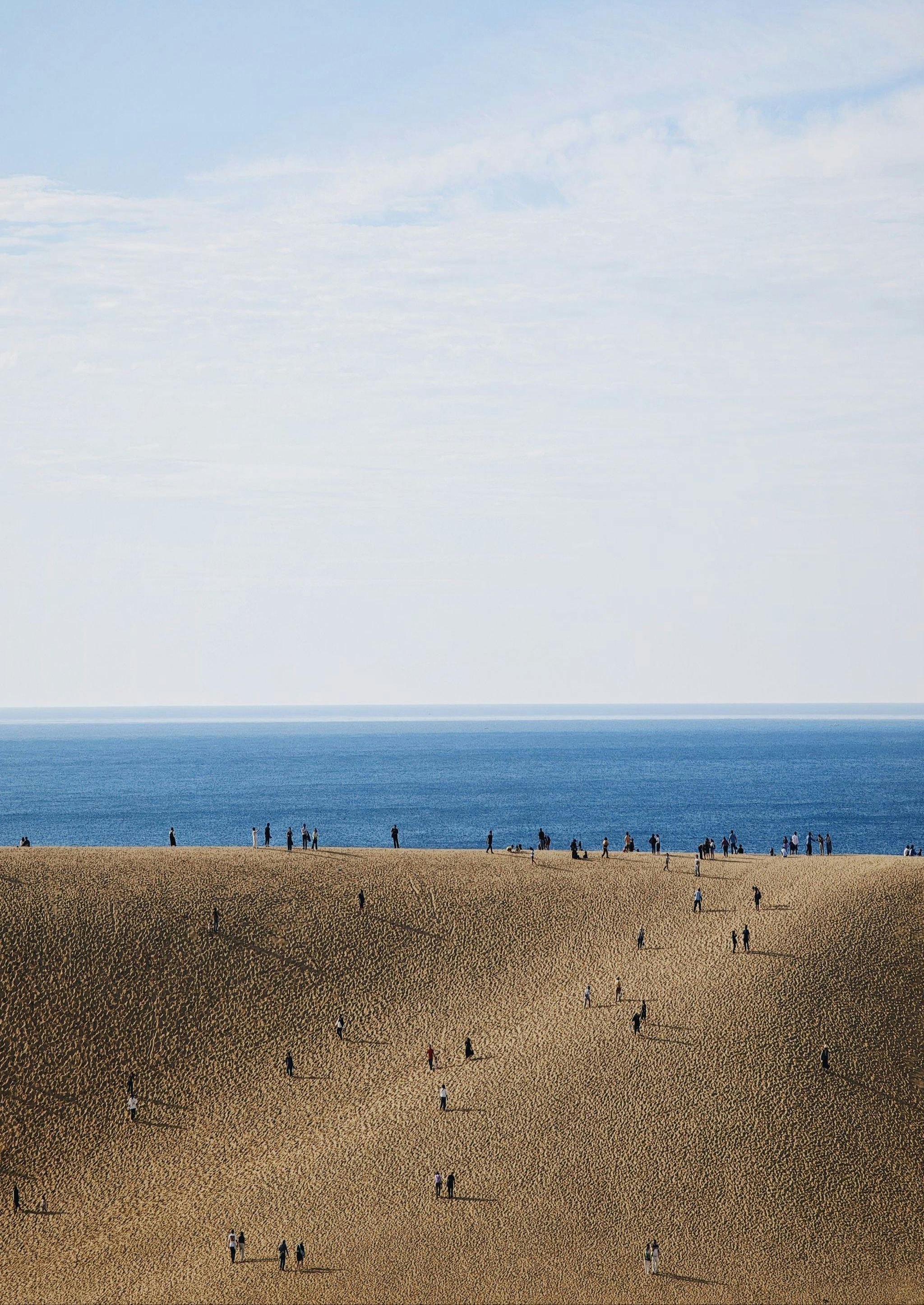}
\definecolor{title}{HTML}{ccddee} 

\makecover

\tableofcontents



\mainmatter

\chapter*{Executive Summary}

This report leverages Principal Component Analysis (PCA) applied to regional price indexes to uncover dominant trends in Australia’s housing market. Regions are assigned PCA-derived scores that show which market forces dominate there, allowing broad local market classifications. This approach highlights where price developments tend to move in unison, even if regions are far apart. The three most dominant trends are described in detail and, along with the scores, provide objective tools for policymakers, researchers, and real estate professionals.

\section*{Insights from PCA}
\begin{itemize}
    \item PCA identifies dominant market trends in Australia. This is extracted from a large collection of local repeat sales price indexes.
    \item Each region receives PCA scores that characterize the local housing market.
    \item The scores are expressed as correlations of local trends to the dominant market trends, and are based on house price data alone, without relying on socioeconomic data.
\end{itemize}

\section*{Dominant Trends by Importance}
\begin{itemize}
    \item \textbf{National Trend:} Reflecting macroeconomic factors, influencing most regions.
    \item \textbf{Mining Trend:} Scores correlate with the mining character of regions, e.g., in WA and QLD, contrasting with negative scores in Sydney.
    \item \textbf{Lifestyle Trend:} Reflecting migration to affordable, high-appeal coastal regions, such as Tea Gardens (NSW) and Surfers Paradise (QLD). Recent migration trends highlight a growing shift toward regional lifestyle regions, reinforcing the PCA-derived lifestyle trend as an important structural housing market factor. This migration pattern, particularly observed since COVID-19, suggests that lifestyle markets are not just influenced by affordability but also by broader demographic shifts driving long-term demand for regional housing spanning decades.
\end{itemize}

\section*{Lifestyle Market Dynamics}
\begin{itemize}
    \item Lifestyle regions lag urban markets like Sydney during booms but catch up sharply near the end.
    \item Despite being in different states, the lifestyle regions in NSW, QLD, and VIC exhibit similar price trends since 1990, marked by two significant booms.
    \item This cyclical pattern challenges the common idea that house prices follow a simple doubling trajectory every decade. Instead, PCA reveals that lifestyle markets experience extended stagnation phases, punctuated by sudden surges tied to urban migration patterns and affordability shifts, catching up with the greater urban areas. 
\end{itemize}

\section*{Mining Market Dynamics}
\begin{itemize}
    \item Mining region prices rose 4–5 times in real terms during the 2000s mining investment boom, peaking in 2012, followed by steep declines.
    \item PCA mining scores identify the most volatile regions, highlighting risks for homeowners and investors, and informing policy decisions.
    \item Resource-intensive states such as QLD, WA, and NT experience extended price stagnations following industry-related shocks. The PCA mining score provides a direct quantitative tool to capture this volatility, distinguishing mining-driven price fluctuations from broader national trends.
\end{itemize}

\section*{Implications}

\textbf{Policymakers and Researchers:}
\begin{itemize}
    \item Improved understanding of urban-rural dynamics and affordability challenges by identifying lifestyle desirability signals.
    \item Enhanced decision-making based on granular, objective insights.
    \item Lifestyle migration from major cities to regional areas has placed significant pressure on housing affordability in high-scoring PCA lifestyle regions. PCA-based insights can help policymakers anticipate where affordability constraints will intensify, particularly in supply-constrained coastal areas.
\end{itemize}

\textbf{Real Estate Professionals:}
\begin{itemize}
    \item Refined strategies for market targeting and advertising based on objective local market knowledge.
    \item Tailored advice for sellers and buyers based on regional trends and characterizing local markets.
\end{itemize}

\textbf{Investors:}
\begin{itemize}
    \item Awareness of cyclical risks and opportunities for informed diversification.
    \item Identification of growth potential and understanding market growth cycles.
    \item The timing of lifestyle booms supports the notion that market spillovers play a key role in driving regional price appreciation. This dynamic challenges the common idea that price trends can be easily anticipated. Instead, PCA shows that lifestyle markets remain cheap relative to urban centers for extended periods before experiencing rapid price surges that can appear sudden but follow cyclical patterns.
\end{itemize}

\section*{Summary}

This report builds on the recent work of  \cite{sijp_francke_ree2024}, applying their methodology in more detail to regional dynamics. By examining granular repeat sales house price indexes for over 2,000 SA2 regions from 1990 to 2024, it identifies lifestyle and mining as major regional price drivers that contrast with Sydney and Melbourne. Principal Component Analysis (PCA) provides objective scores that describe local housing market behavior in terms of these factors.

PCA is a statistical method that identifies patterns in large datasets by breaking down complex trends into a smaller number of key components. These components explain shared movements in price indexes across regions, revealing market forces at play. The PCA-derived scores highlight dominant market trends and cyclical growth patterns, such as the national, mining, and lifestyle components, while providing actionable insights for stakeholders.

By linking housing price trends to broader economic phenomena, this analysis offers tools for policymakers, real estate professionals, and investors to navigate the housing landscape. Policymakers can better understand urban-rural dynamics and affordability, real estate businesses can refine their strategies, for instance in marketing, using objective market characterizations, and investors can make more informed decisions about market cycles and diversification.

\chapter{Framework}

\section{Introduction}

Understanding the distinct characteristics of local housing markets is important for analyzing regional economic dynamics and making informed policy or investment decisions. However, conventional property market indexes often lack the granularity needed to capture the complexities of mixed and diverse housing markets, especially in regions facing affordability pressures, economic shifts, or the interplay between urban and regional dynamics. Moreover, objective, data-driven measures to classify markets systematically and quantify their key drivers remain limited.

This report addresses these gaps by applying Principal Component Analysis (PCA) to local price indexes in Australia. PCA is a statistical technique used to identify dominant trends in complex datasets by reducing dimensionality while preserving most of the variance. In the atmospheric sciences, this is widely used to extract key spatial patterns that evolve in time from geographically distributed datasets, for instance temperature time series at various locations. It therefore has a long-standing history in geographic time series settings and is also straightforward to implement.

Here, it is used similarly, by leveraging geographically distributed granular repeat sales indexes, summarizing the price behavior for a large collection of regions in terms of a small set of trends. Based on the results, this analysis introduces lifestyle and mining scores, data-driven metrics that systematically quantify the key forces shaping regional housing trends. 
In turn, by expressing local price index in terms of the main trends, PCA provides a way of smoothing to the price indexes that is informed by local and national factors.

The PCA-derived scores are the central contribution of this report, providing a novel framework for understanding local housing market behavior. The analysis puts these scores (obtained from the price data) in perspective and characterizes their meaning by linking them to key regional dynamics, including lifestyle migration from the cities and the mining boom. These patterns shape the geography of the Australian housing markets and are therefore important in classifying the local markets.
While these local insights are valuable, the scores themselves have broader applications for decision-making and analysis.

Lifestyle markets are regions celebrated for their natural beauty, recreational opportunities, and appeal as destinations for second-home ownership. These same factors also attract tourism. Top-scoring regions like Batemans Bay, Forster, Noosa and Surfers Paradise are recognizable examples. These areas often attract urban buyers seeking affordability, with an emphasis on a better quality of life through proximity to beaches, scenic landscapes, and a slower pace of living. Alternatively, tourism and second home ownership may put their stamp on a local lifestyle housing market. PCA identifies a dominant lifestyle-driven trend (PC~3) that captures the unique dynamics of these markets, distinct from the national trend (PC~1) and the mining-driven trend (PC~2). Importantly, these trends are entirely data-driven, with no reliance on predefined lifestyle factors.

Similarly, the mineral boom from the early 2000s to the mid-2010s profoundly shaped resource-rich housing markets. PCA captures this phenomenon in a mining-driven trend (PC~2), which aligns with regional economic developments and contrasts with the urban markets of the eastern states, particularly Sydney and to a lesser extent Melbourne, which exhibit negative scores.

While Australia’s main property market indexes reasonably track price changes, they often fail to reflect finer-grained trends that are essential for understanding local market dynamics. This report overcomes these limitations by producing high-quality repeat sales indexes at the SA2 regional level. Together with PCA, these methods provide a powerful framework for smoothing local indexes and revealing hidden patterns in housing market behavior.

This report demonstrates the practical utility of PCA-derived scores. It enables the identification of similar submarkets, even if they are many kilometers apart. This sheds fresh light on the behavior of local house price indexes, as they are no longer viewed in isolation.

\section{Methodology}

This report applies the methods of  \cite{sijp_francke_ree2024}. Local house price trends are calculated for each SA2 region using a repeat sales method as a starting point, which tracks price changes over time for properties that have sold more than once. To ensure smoother and more reliable results, the trends are adjusted by considering how neighboring regions influence price growth. PCA (Principal Component Analysis) is then used to analyze these trends and refine the indexes. This technique helps identify patterns and relationships between markets, regardless of their geographical separation. This section provides a brief overview of the approach. For more detailed mathematical formulations of the repeat sales method and PCA, refer to Appendix \ref{method_detail} and \cite{sijp_francke_ree2024}.

\subsection{Local price indexes}
The methodology begins by constructing granular repeat sales indexes for each of around 2200 SA2 regions in Australia. SA2 regions are medium-sized statistical areas designed to reflect communities that interact socially and economically, typically with populations between 3,000 and 25,000 people. Fitting price indexes separately for each SA2 region would lead to excessive volatility due to sparse data in many areas. To mitigate this, the regression model jointly estimates all local indexes within a (larger) SA4 region while enforcing smoothness across time and space. SA4 regions are broader geographic units used for labor force analysis, each typically containing around 25 SA2 regions.

\begin{wrapfigure}{r}{0.4\textwidth} 
    \centering
\includegraphics[width=0.3\textwidth]{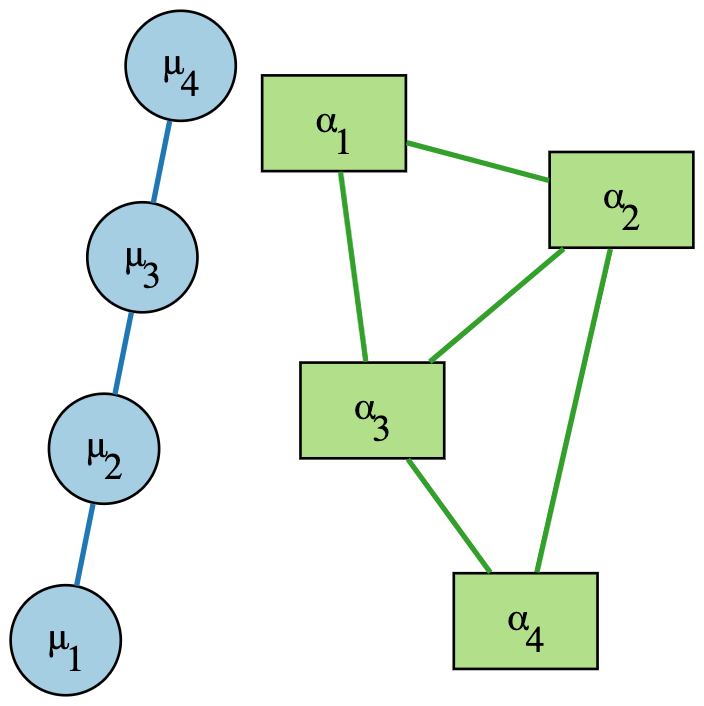}
\caption{Schematic representation of using graphs to model relationships among regions or time. The blue network on the left represents a path that could model time, whereas the green represents adjacent regions.}
\label{fig:cartoon}
\end{wrapfigure}

Volatility is mitigated by encouraging local index values to resemble those of nearby locations and times. To capture spatial dependencies, the model represents regions as nodes in a network, such as shown schematically in Figure~\ref{fig:cartoon}, where connections between nodes define neighboring relationships. Similarly, time is represented as a linear network, with connections between subsequent months. Differences in index values along these connections are mitigated by assuming these gradients follow a centered Gaussian distribution, imposing a penalty on large deviations so that trends evolve smoothly.

The regularization is essentially a generalized ridge regression in transformed coordinates, penalizing price differences between locations and times. This prevents excessive fluctuations and naturally imputes missing values by borrowing information from nearby data points. The result is a set of stabilized, regionally consistent price indexes that can be used for further analysis.

\subsection{PCA}

The second step is PCA applied to the set of regional price indexes  obtained from the first step. This identifies the main uncorrelated underlying trends (time series) in the Australian market. 
PCA also yields a score for each region, showing how strongly it is influenced by each main trend.

PCA is a statistical technique that is used for interpreting large datasets.
The main purpose of PCA is to simplify a dataset with many interrelated variables while preserving as much of its variability as possible. This is done by creating a new set of uncorrelated variables, called principal components (PCs), which are arranged in order of importance so that the first few capture the majority of the variability in the original data.
Here, the variables are the house price index value for each region, and the PCs are the PC trends (shown in figure \ref{pcs} of the Appendix).

The procedure is to first find a new time series, similar to a ``price index'', that is a weighted sum of all the regional trends and has maximum variance: this yields the most dominant trend. The next step is to do the same by finding the most dominant trend as a weighted sum of the regional trends, provided that that trend is uncorrelated with the first dominant trend already found. This yields the second PC trend, and so on. This procedure is ultimately based on the mutual correlations between the price indexes of the regions, and therefore provides information on where house prices move in unison.

As is common in data that have not been detrended, the first house price index trend found in this study explains 92\% of the variance. Around 95\% of the variance is explained by the first two PC trends and around 96.5\% for the first six. This means that much of the dataset can be reconstructed using weighted sums of the first few PC trends alone as, conversely, each local regional trend can be calculated as a weighted sum of the PC trends. The associated weights are used in this report as a score indicating how strongly each region is influenced by each PC trend. These weights can also be calculated as correlations to those main trends, as is done in this report.

For instance, a "lifestyle" trend is identified: the weight coupling a region to this trend is used as a "lifestyle score" for that region. Examples in NSW of areas scoring high on the lifestyle scale are Tea Gardens, north of Newcastle, and Batemans Bay in Southern coastal NSW score. This is solely based on the price index behavior of the local housing market and their correlations to other regions, so it is important to keep in mind that these scores are not based on particular direct desirability criteria or recent growth ``performance''.

\section{Data}\label{sec:Data}

The methods described here are applied to a compiled dataset of around three million sales records  originating from state valuer general data. Included are transaction prices (in Australian dollars), dates of sale and latitude and longitude of house location. We examine houses and not apartments. 
We examine house price changes in the Australian Statistical Areas Level~2 (SA2) regions defined in the Australian Statistical Geography Standard (ASGS) used by the Australian Bureau of Statistics (ABS) and described in \cite{abs:2021}. Statistical Areas Level 2 are medium-sized regions that represent a community that interacts together socially and economically. Equivalent to around 2.5 US census tracts, they cover Australia without overlap and have a population range of 3,000 to 25,000 persons and an average population of about 10,000 persons. We use 2052 different regions across Australia in this study.
Each SA2 region has around 2-5 sales per month in the dataset. The monthly availability of at least one recorded house sale at the granular SA2 region level is 85\% nationally.

\chapter{Lifestyle Markets: Identifying Sea Change Booms with PCA}

Lifestyle is the third most dominant trend revealed by the PCA. As it displays rich behavior, it is presented first.
Based on geographical evidence and a strong correlation of the PC 3 scores with lifestyle word frequencies in listings data, this section will provide support for this interpretation and describe these markets in more detail. 
Lifestyle  regions are found in all states and territories except NT.

\section{Geography of the lifestyle regions}

Examining which regions score high on PC 3 reveals that this score must reflect lifestyle.
The top PC 3 scoring SA2 regions for each major city statistical area (table~\ref{table:eof3} in the Appendix) are virtually all coastal regions with a beach lifestyle that are often also popular for second homeownership. Examples are Tea Gardens, Forster - Tuncurry region, Batemans Bay and Ulladulla in Rest of NSW, Port Douglas, Daintree and Surfers Paradise in Rest of QLD, Budgewoi - Buff Point - Halekulani, Warnervale, The Entrance in Sydney and Redland Islands in Brisbane.

These regions also attract visitors from the urban regions, and some even attract major international tourism. This, along with closer analysis of the high-scoring regions, indicates that PC 3 is associated with lifestyle. It is important to realize that high scores here means having the strongest PC 3 characteristic, not an initial value judgement entered into the model by the researcher.

Based on this geography and findings below, a positive lifestyle score indicates that there is  a lifestyle element in that local market to some degree, and the stronger the score, the more this element dominates. The regions with a lifestyle score of  $0.1$ or above are clearly regions with a lifestyle appeal, although there are many regions with scores between $0.05$ and $0.1$ that are still well known for lifestyle. Scores above $0.25$ indicate the solid and well known coastal lifestyle regions.

The top scoring regions in Table~\ref{table:eof3} (Appendix) suggest two types of lifestyle regions. First, tourism-driven regions, areas where the lifestyle appeal primarily supports short-term visitors, second-home ownership, and holiday accommodations. Examples include Port Douglas (QLD) and Kangaroo Island (SA), which attract major tourism and seasonal residents but have relatively fewer permanent lifestyle migrants. The local economies in these regions often depend on tourism, and are less characterized by a large influx of full-time residents.
the second type of lifestyle region involves permanent lifestyle migration. These are the regions that attract long-term residents seeking a sea change or tree change (terms used in Australia to indicate migration to scenic regions), often supported by infrastructure for remote work, retirement migration, or commuting access to major cities. Examples include Surfers Paradise (QLD) and the Central Coast (NSW), which have both strong lifestyle appeal and substantial residential growth. Many of these regions have experienced increased migration due to flexible work arrangements post-pandemic. It is interesting that both types of regions share broad growth characteristics.

\begin{wrapfigure}{l}{0.7\textwidth} 
    \centering
\includegraphics[width=0.6\textwidth]{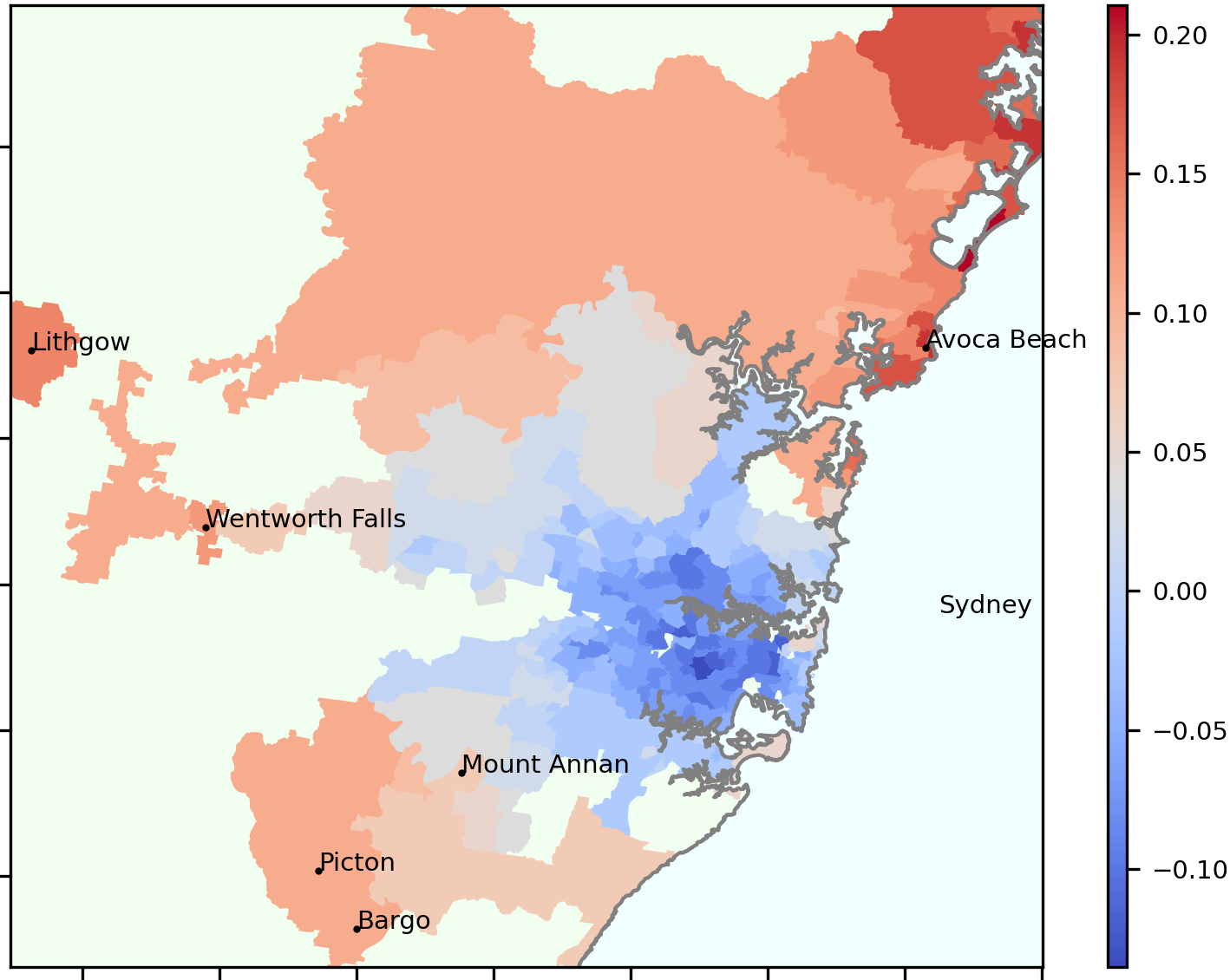}
\caption{PC 3 scores for Greater Sydney, associated with lifestyle. The color of each SA2 region indicate its PC 3 score. Analysis reveals that this score measures lifestyle. The negative values within Sydney, and the positive values in the lifestyle regions, particularly to the northeast, reflect the relationship between inner-city suburbs versus more affordable outer city and regional areas possessing considerable lifestyle or ``holiday'' appeal, involving periods of population movement and second home buying away from the inner city.     }
\label{map_sydney_eof2}
\end{wrapfigure}

In NSW, the top PC~3 regions are located immediately north of Newcastle and along the south coast near Batemans Bay. Within the Greater Sydney region more specifically, the top regions are located within the Central Coast, over an hour of driving time from the inner city and characterized by beaches, surf and holiday homes.
The high PC 3 lifestyle scores are not found in inner Sydney, where scores are negative (figure~\ref{map_sydney_eof2}).

The high positive score for the Central Coast, e.g. Avoca Beach, is visible to the north (again, figure~\ref{map_sydney_eof2}). 
Of interest is the high PC 3 (lifestyle) score along the narrow ribbon of towns situated along a narrow plateau in the scenic Blue Mountains to the west of Sydney, increasing along the Great Western Highway going west, past Wentworth Falls and to Lithgow, the beginning of the lower lying Central West.  
Away from Sydney inland to the southwest, more positive scores begins at Mount Annan, where the city meets the rural countryside of regional New South Wales. These are examples of lifestyle areas in scenic non-urban areas \emph{away} from the coast.
The other largest cities show a similar pattern (figure not shown).

While the highest-scoring lifestyle regions in New South Wales are coastal areas known for their beaches, there are other regional markets away from the ocean with a significant, albeit lower, lifestyle score. These include Griffith, a vibrant agricultural hub renowned for its rich produce and wineries and rural charm. Similarly, Wellington, located in the picturesque Central Western Slopes region of New South Wales, combines scenic landscapes with a slower pace of life. These regions highlight the diverse ways in which lifestyle attributes manifest across the state, extending beyond the traditional coastal paradigms.

\begin{wrapfigure}{l}{0.7\textwidth} 
    \centering
\includegraphics[width=0.6\textwidth]{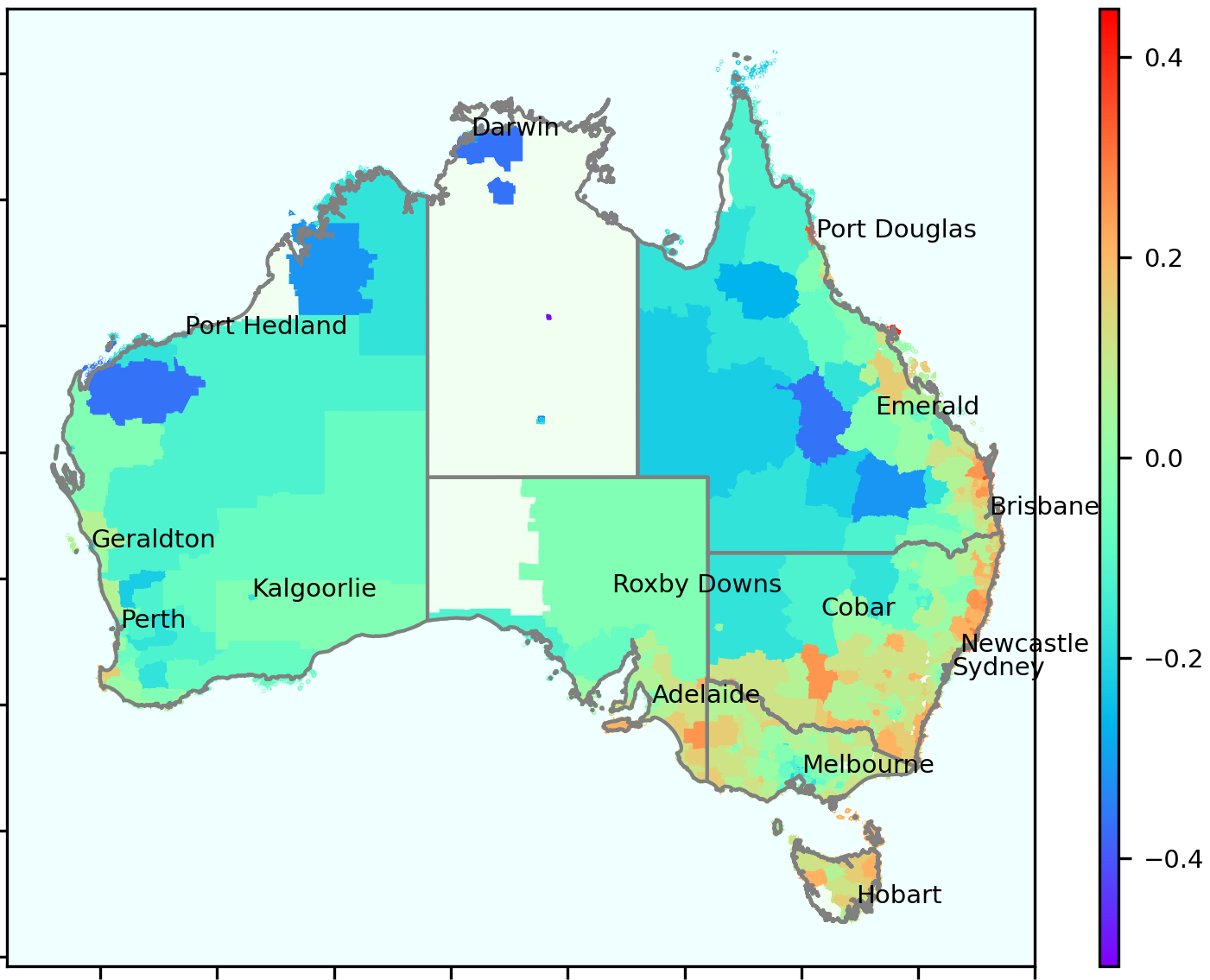}
\caption{PC 3 scores for Australia, associated with lifestyle. The color of each SA2 region reflects its PC 3 score. }
\label{map_au_lifestyle}
\end{wrapfigure}

A broader national overview for the lifestyle score is shown in figure~\ref{map_au_lifestyle}.  The lifestyle scores are high on the east coast, but there are also coastal lifestyle regions in WA concentrated near Mandurah and down to Margaret River in the southwest, albeit with somewhat lower score.
The scenic regions around Mount Gambier near the VIC border, and Victor Harbour and Kangaroo Island in SA also stand out in the figure. Darwin and many regions in the interior of the continent have negative PC 3 lifestyle scores. 
There are high scoring regions such as Port Douglas that are barely visible on the map due to size, and perhaps better examined in table~\ref{table:eof3} (Appendix). The highly urbanized regions with negative PC 3 scores are also almost too small to see at this scale, except for Melbourne.

\section{Evidence for lifestyle from listing advertisements}
\label{wordcount}

The lifestyle character of the PC 3 score is also apparent from relative word frequencies in property listings data for 1990 to present. Based on determining the relative word frequencies in house-for-sale advertisements found in listings data, spatial correlations with the PC 3 score (by SA2 region) are determined. The strongest spatial correlation (over the SA2 regions) between the frequency of a particular word and the PC 3 score is for the words holiday, little, view, acre, coastal, package, sea, value, neat, rural, affordable, ocean, cottage, country, pin, close, beach, gem, life and paradise. The correlations are between 0.32 and 0.47. 

It is immediately clear that many of the top scoring words relate to lifestyle. The word ``affordable'' also scores high, reflecting that indeed affordability is part of the attraction of the lifestyle regions. High correlation words that almost made it into the table were ``surf'' and the word ``lifestyle'' itself. The same analysis for PC 1 and PC 2 does not yield any correlations to lifestyle words, indicating that PC 3 is the main indicator for this phenomenon.

\begin{wrapfigure}{r}{0.7\textwidth} 
    \centering
\includegraphics[width=0.6\textwidth]{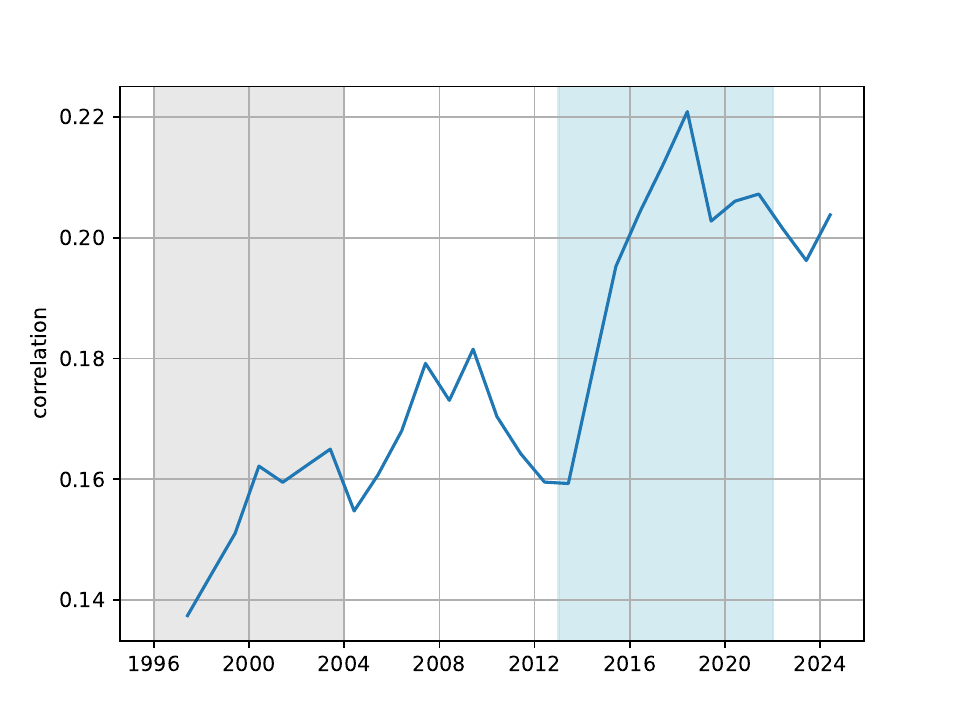}
\caption{Correlation between the frequency of lifestyle words for each year and the PC lifestyle indicator. }
\label{word_correlation}
\end{wrapfigure}

The word-group averaged correlation between the relative frequencies of a collection of lifestyle words tracked over time shows that the association between the use of lifestyle words in property listing ads and the lifestyle regions increased over the past few decades (figure~\ref{word_correlation}).

As the agent's choice of words likely reflects her perception of what is considered desirable among buyers, these word counts provide further evidence that PC 3 is indeed related to lifestyle. It also shows a rising emphasis on these characteristics in the lifestyle region markets when compared to other markets, and it is intriguing that this increase took place during the two boom periods described in the following (compare the periods of increase with the shaded areas in figure~\ref{word_correlation}).

\clearpage
\section{Price developments in the city lead the lifestyle markets}

The lifestyle housing markets have experienced two significant booms since 1990, both driven by migration from inner cities, where the booms begin more gradually, end with a sharp price increase, and are interrupted by periods of stagnation. The most recent boom has further exacerbated the current affordability crisis.

\subsection{The lifestyle markets since 1990}

\begin{wrapfigure}{l}{0.7\textwidth} 
    \centering
\includegraphics[width=0.6\textwidth]{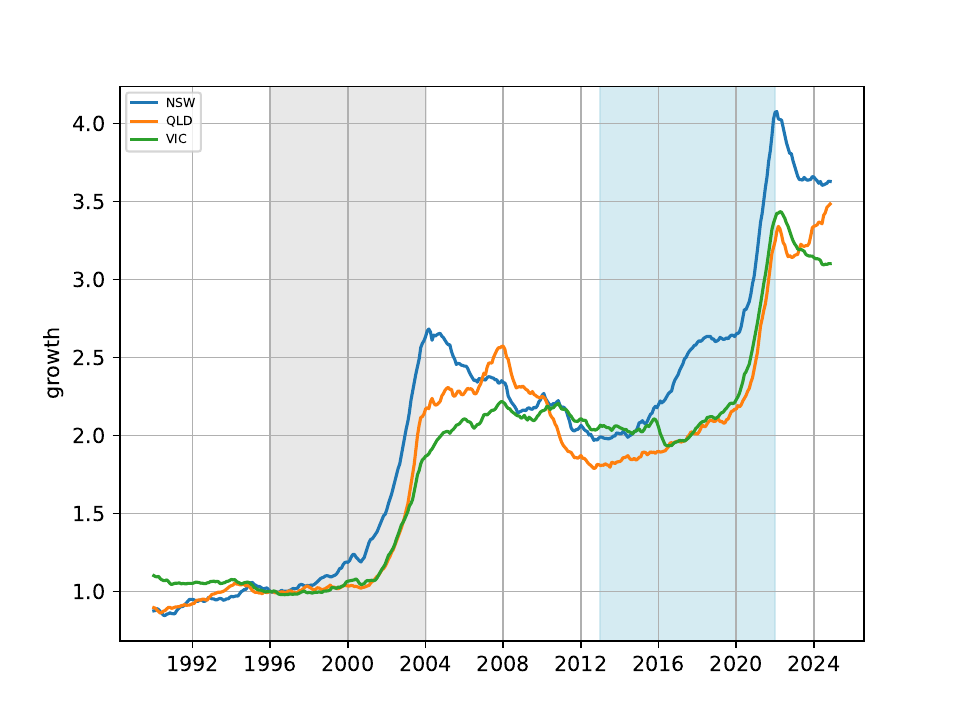}
\caption{CPI-corrected price index averaged over the top 20 lifestyle regions for three states. The indexes are normalized at Jan 1996, the beginning of the first boom. The shaded areas indicate the boom phases of the Sydney market, see below. There are two lifestyle boom phases, where the first is followed by a long period of stagnation, and even decline in real terms.}
\label{states_lifestyle}
\end{wrapfigure}

The lifestyle housing markets can be tracked by taking the top 20 lifestyle SA2 regions\footnote{For NSW: Tea Gardens - Hawks Nest, Nelson Bay Peninsula, Forster, Tuncurry, Nambucca Heads, Batemans Bay - South, Forster - Tuncurry Region, Old Bar - Manning Point - Red Head, South West Rocks, Sussex Inlet - Berrara, Anna Bay, Broulee - Tomakin, Callala Bay - Currarong, Terranora - North Tumblegum, Ulladulla Region, Kempsey Region, Batemans Bay, Moruya - Tuross Head, Taree Region, Eden. \\ For QLD: Airlie - Whitsundays, Main Beach, Redland Islands, Peregian Springs, Daintree, Cooloola, Port Douglas, Booral - River Heads, Magnetic Island, Surfers Paradise, Sunshine Beach, Noosa Heads, Torquay - Scarness - Kawungan, Craignish - Dundowran Beach, Noosaville, Urangan - Wondunna, Caloundra - Kings Beach, Buddina - Minyama, Peregian Beach - Marcus Beach, Granville \\
For VIC: Robinvale, Glenelg (vic.), Merbein, Hamilton (vic.), Portland, Gannawarra, Red Cliffs, Nhill Region, Corangamite - North, Morwell, Yarram, West Wodonga, Southern Grampians, Corangamite - South, Orbost, Longford - Loch Sport, Lakes Entrance, Moyne - East, Chiltern - Indigo Valley, West Wimmera} according to PC 3 score in each state and combine their local indexes into a weighted mean index. 
The house price index reveals remarkably similar growth patterns for NSW, QLD, and VIC since 1990 (figure~\ref{states_lifestyle}). These markets have experienced two distinct boom periods in real terms, each preceded by a period of stagnation. The first boom is linked to the well-known surge starting around 1996–1997 in NSW and ending for all three states in 2004 \cite[e.g., as examined for Sydney by][]{waltl2019}. The lifestyle regions began to boom around 2001, starting somewhat later in QLD and VIC. The second boom saw the sharpest increase in prices from 2020 to 2022, leaving prices at elevated levels after that, with no further increases.

These boom periods, while substantial, were followed by significant stagnation, highlighting the more cyclical nature of lifestyle market growth. For instance, the NSW lifestyle market lost 34\% of its value in real terms after the first boom, only to recover its 2004 peak by July 2018. The QLD market lost 21\% and recovered its 2004 levels in 2020, with a similar stagnation period for VIC. These periods of stagnation post-2004 echo the earlier decade of low or absent real growth in the 1990s preceding the first boom.

Although Sydney is among other urban centers that export sea changers, as this is intended to be a brief report, Sydney will serve as the example of the large metropolitan regions to study in relation to the lifestyle market limited to NSW. The other states also receive a significant influx of Sydneysiders, particularly QLD. The lifestyle regions in these other states show roughly similar behavior (see figure~\ref{states_lifestyle}). 
For a detailed analysis of the migration patterns to and from major urban areas and regional cities, including interstate, and their effect on house prices, we refer to \cite{yanotti2023}.

\begin{wrapfigure}{r}{0.7\textwidth} 
    \centering
\includegraphics[width=0.6\textwidth]{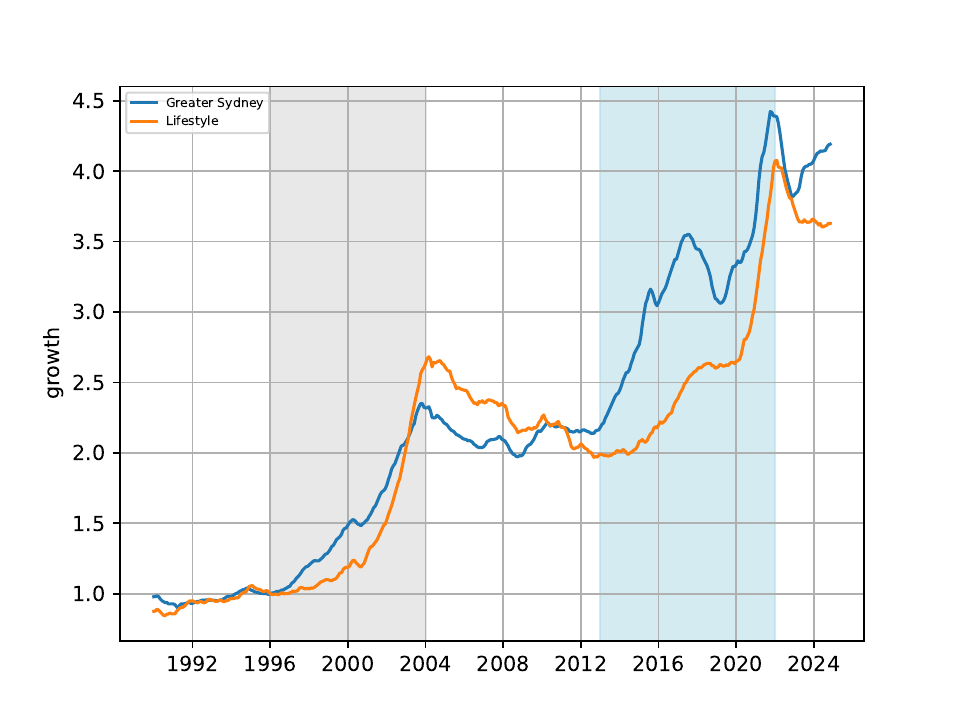}
\caption{CPI-corrected price index for Sydney and the index averaged over the top 20 lifestyle regions in NSW. The indexes are normalized at Jan 1996, the beginning of the first boom. The shaded areas indicate the boom phases of the Sydney market. Growth is more steady in Sydney, and the  lifestyle markets catch up to each Sydney boom with a steep price increase after a delay.}
\label{city_lifestyle}
\end{wrapfigure}

\begin{wrapfigure}{l}{0.7\textwidth} 
    \centering
\includegraphics[width=0.6\textwidth]{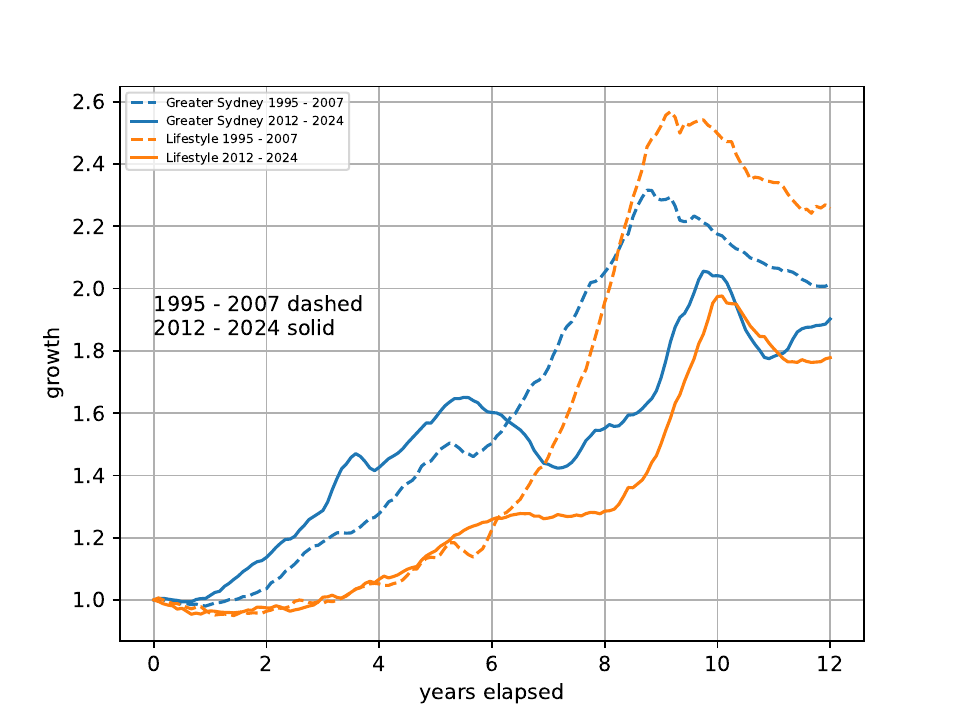}
\caption{CPI-corrected price index for Sydney and the top 20 NSW lifestyle regions combined for the two growth periods overlaid. The horizontal axis denotes years elapsed since 1995 for the first period (dashed) and 2012 for the second (solid). Compare the blue to the orange curve for the dashed lines for the first boom, and the solid lines for the second boom. The two booms have been roughly similar in size, and unfolded similarly, with the lifestyle markets initially lagging and then catching up with a sharp increase.   }
\label{city_lifestyle_diff}
\end{wrapfigure}

Since 1996, the Sydney housing market and the lifestyle market in NSW have shown similar real price growth over the period 1990 to present (figure~\ref{city_lifestyle}). However, Sydney's price growth has been steadier, with a milder decline after 2004. Comparing the two boom periods by overlaying them on a single timescale, as in figure~\ref{city_lifestyle_diff}, reveals similarities. 
During both periods, house prices approximately doubled in Sydney and also the lifestyle regions. However, the first boom was larger and somewhat shorter, with lifestyle markets peaking at nearly 2.6 times their initial value, and around 2.3 times in Sydney.

Both booms took on the order of a decade to peak, and the lifestyle market peaked very shortly after Sydney during both booms.
 A price gap develops in the earlier phases of each boom, where the NSW lifestyle market becomes increasingly cheap compared to Sydney. 
 A  delayed and sharper surge in lifestyle markets then closes the price gap that has developed between the two. During these booms, lifestyle markets, initially appearing more affordable compared to Sydney than they previously were, catch up sharply. This "catching-up" phase is characterized by shorter, sharper growth periods, such as 2001–2004 during the first boom and 2020–2022 during the second.

\subsection{City equity enables a sea change }

The relative frequency of lifestyle words in listing advertisements became more strongly correlated with the lifestyle regions in NSW during the second boom, remaining high afterward (see figure~\ref{word_correlation}, and compare to figure~\ref{city_lifestyle}). This highlights the growing importance of lifestyle aspects for buyers in periods of high valuations, along with the perceived affordability of these destinations. House prices are well-established drivers of migration \cite[e.g.][]{crommelin2022}, which, in turn, fuels local demand in destination regions.

 There are sharp price increases that eliminate the lag in lifestyle markets at the end of each boom (see figure~\ref{city_lifestyle_diff}).
This likely results from urban migration to lifestyle regions, as documented more broadly for the recent period by \cite{yanotti2023}. This delayed and eventually more rapid catching up may give the impression of a particularly strong lifestyle boom during the final two years of each period. This sharp rise indicates the effect of recent inner city price gains working through in these markets. The delay may arise from the time it takes to accumulate equity and for the deepening of the price differential, increasing the probability for an inner-city household to make a decision for a sea-change or the purchase of a second home. If so, the timing of these lifestyle booms may involve longer-term socio-economic dynamics than is often cited in the media.

Also, the booms support the notion that market spillovers play a key role in driving regional price appreciation. This dynamic challenges the common market myth that price trends can be easily anticipated (market timing). Instead, PCA shows that lifestyle markets remain cheap relative to urban centers for extended periods before experiencing more rapid price surges.

This report does not explore broader determinants of aggregate Australian housing prices. For example, \cite{wang2018} identify key factors such as mortgage interest rates, consumer sentiment, the S\&P/ASX 200 stock market index, and unemployment rates. Similarly, international immigration contributes to housing demand, as noted by \cite{moallemi2020}, particularly in Sydney and Melbourne. The common national PCA trend is likely to be influenced by these factors.

There is a significant relationship between house price changes and population movements \cite{meen1996}. Lifestyle market prices are shaped by both local demand and an inflow of urban buyers, with the latter emerging as the dominant force based on the magnitude and timing of price increases discussed here. These findings align with \cite{yanotti2023}, who emphasize that migration from urban centers to regional areas plays an increasingly important role in Australian house price dynamics. However, these destination regions, historically valued for affordability, are now home to many low-income and tenant households facing mounting housing cost pressures.

The high-lifestyle-score regions identified in this report have some overlap with the migration destinations described by \cite{yanotti2023}, particularly on the basis of affordability, but are more narrowly focused. While \cite{yanotti2023} examine a broad range of areas away from the inner city, here the PCA lifestyle score specifically highlights regions characterized by attractions such as natural beauty and recreational appeal, often associated with the ocean. 

Adding to the house price pressures arising from these desirable aspects is that such regions are the obvious locations for a holiday home, to be rented out to tourists or used as a second home. 
Many high-scoring PCA lifestyle regions also face structural supply constraints, such as strict zoning regulations and environmental protections that limit new housing development. These constraints exacerbate price surges when migration-driven demand increases, particularly in coastal areas.

The negative lifestyle scores in inner Sydney (see figure~\ref{map_sydney_eof2}) are interesting, and suggestive of the push from the inner city to the lifestyle regions.
Sydney comprises a large urban area of varying density, characterized by a sprawling metropolis, significant population pressures, and congestion challenges. 
The PC score itself does not directly identify any underlying migration flows that may cause the price dynamics under study.  Instead, the migration patterns described by \cite{yanotti2023} are cited here to explain the scores and price developments in the lifestyle markets.

Placing these dynamics between Sydney and the NSW lifestyle regions within the context of the two major house price boom cycles over the past 34 years, this report demonstrates that similar patterns were already evident around 2004. This underscores the persistent and cyclical nature of these trends.
\cite{yanotti2023} describe significant price impacts caused by the intensification of migration from inner-city suburbs to outer city and regional areas during the COVID-19 pandemic, suggesting that the dynamics described here are unlikely to abate in the near future.

\chapter{Mining and Housing: Insights from the second largest PCA trend}
\label{sec:mode2}

This section examines the region-aggregated scores for PC 1 and 2 and attempts to identify general factors relating to the PC 2 trend, along with some regional examples. 
The evidence that PC 2 represents mining is based on geography and a strong correlation of the PC 2 scores with mining employment.

\section{The resources industry drives PC trend 2}

Examining which regions score high on PC 2 reveals that this score must reflect the effect of the mining industry on local housing markets. 
Table~\ref{table:city_corr} shows mayor-city means of scores of PC 1 and 2.  
The strongest positive influence of PC 2 is in mining-dominated WA, followed by regions known to be influenced by the resources sector. This geographic pattern suggests that positive values of the PC 2 score indicate an influence of mining, as will be confirmed below. Sydney has a strong negative score of -0.25 and Melbourne a more weakly negative score of -0.08, whereas Brisbane has a neutral score. The overall score for the other main trend, PC 1, is high for all areas, with the rest of WA and the NT somewhat weaker. This is in agreement with its interpretation as the main national trend.

\noindent  
\begin{minipage}{0.45\textwidth}  
    \centering
    \captionsetup{type=figure}
    \includegraphics[width=\textwidth]{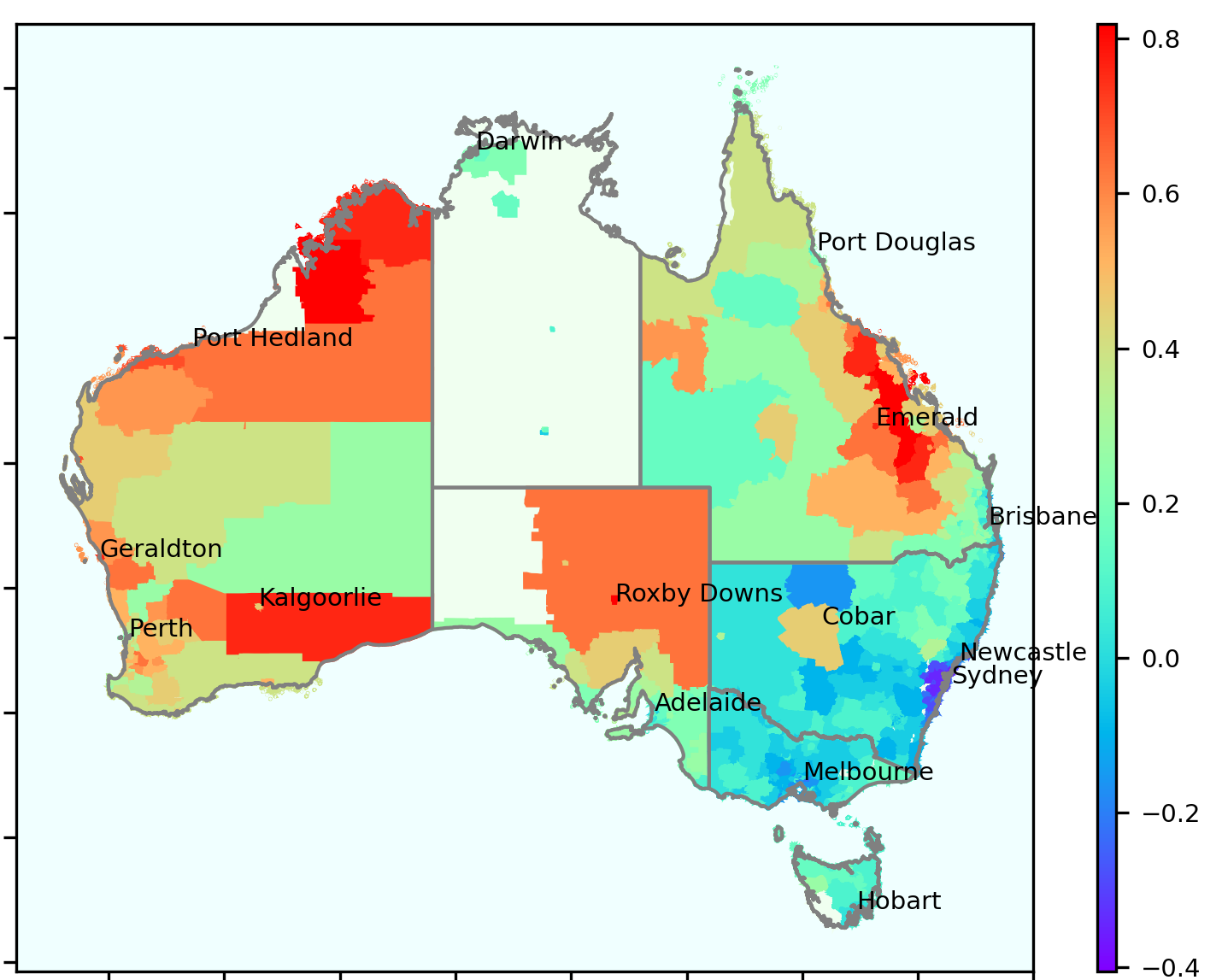}  
    \captionof{figure}{PC 2 scores are associated with mining. The color of each SA2 region reflects its PC 2 score. Analysis reveals that this score measures mining. Of particular interest is the red region near Emerald, QLD: this reflects the shape of the Bowen Basin, an area rich in coal. Although not easily visible here, the Tasmanian west coast also has regions with a moderately positive score reflecting mining.}
    \label{eof1}
\end{minipage}
\hfill  
\begin{minipage}{0.45\textwidth}
    \centering
    \begin{tabular}{lrr}
    \toprule
     Major city &   PC 1 &  PC 2 \\
    \midrule
     Rest of WA &              0.83 &      0.48 \\
     Rest of NT &              0.39 &      0.36 \\
          Perth &              0.93 &      0.33 \\
     Rest of SA &              0.92 &      0.26 \\
         Darwin &              0.88 &      0.25 \\
    Rest of QLD &              0.91 &      0.24 \\
    Rest of TAS &              0.96 &      0.14 \\
       Brisbane &              0.97 &      0.12 \\
       Adelaide &              0.98 &      0.11 \\
    Rest of VIC &              0.98 &      0.03 \\
         Hobart &              0.98 &      0.03 \\
            ACT &              0.97 &     -0.03 \\
    Rest of NSW &              0.96 &     -0.04 \\
      Melbourne &              0.99 &     -0.08 \\
         Sydney &              0.95 &     -0.26 \\
    \bottomrule
    \end{tabular}
\captionsetup{type=figure}
    \captionof{table}{Geographical mean of PC 1 and 2 (mining) scores for major city statistical areas. Means are weighted by house sales volume per SA2 region. 
        The rows are sorted by PC~2 in descending order. }
    \label{table:city_corr}
\end{minipage}

The mining connection is confirmed by external economic data, as table~\ref{table:abs_corr} shows that there is a strong and positive correlation between PC 2 and the fraction of employment in the mining sector per SA2 region, with a correlation of 0.70,
whereas PC 1, the national trend, has a negative correlation of -0.41, and the other PCs a relatively small correlation. 
This shows that local housing markets in regions where the resources sector exerts a significant effect on the local economy are primarily characterized by high positive PC~2 scores. They also have a weaker coupling to the main trend 1.

Despite its urban character, Greater Perth exhibits a positive PC 2 score with an average of 0.33, reflecting the significant influence of the mining industry on its housing market (again, table~\ref{table:city_corr}). This aligns with the well-established understanding that the resources sector contributes substantially to Perth's economy. However, the city's housing market also demonstrates a stronger coupling to the national trend (PC 1) compared to the rest of WA. This shows that, although strongly influenced by the resources sector, the Perth housing market is yet distinct from the regional towns where mining actually takes place.

\begin{table}[ht]
	\centering
	
\begin{tabular}{lrrrrrr}
\toprule
Score &   1 &    2 &    3 &     4 &     5 &     6 \\
\midrule
Corr & -0.41 & 0.70 & 0.22 & -0.01 & -0.13 & -0.08 \\
\bottomrule
\end{tabular}

	\caption{Correlation between fraction of employment in mining sector data and the scores.}
	\label{table:abs_corr}
     \footnotesize
    \begin{tablenotes}
       \item \textit{Notes}: Spearman rank correlation coefficient between 2016 ABS Census SA2-level fraction of employment in mining sector data and scores 1 to 6. This is a spatial correlation using all SA2 regions.
    \end{tablenotes}

\end{table}

Figure~\ref{eof1} shows a map of the PC 2 scores per SA2 region (the mining score). 
Regions known to be associated with the Australian resources sector have high (positive) scores, e.g. the Pilbara in northern WA, Kalgoorlie in southern WA, Roxby Downs in SA and the areas around the coal-rich Bowen Basin in Queensland: in fact, the rough outline of this geological formations is visible in the map.  In contrast, the large eastern urban centers of Sydney and Melbourne, areas of strong non-mining related economic activity, have negative (PC 2) correlations. 
Based on Table~\ref{table:city_corr}, regions within NSW that have a relatively negative PC~2 could reasonably be expected to have similarities to Sydney, and in Victoria to Melbourne.

\begin{wrapfigure}{l}{0.7\textwidth} 
    \centering
\includegraphics[width=0.6\textwidth]{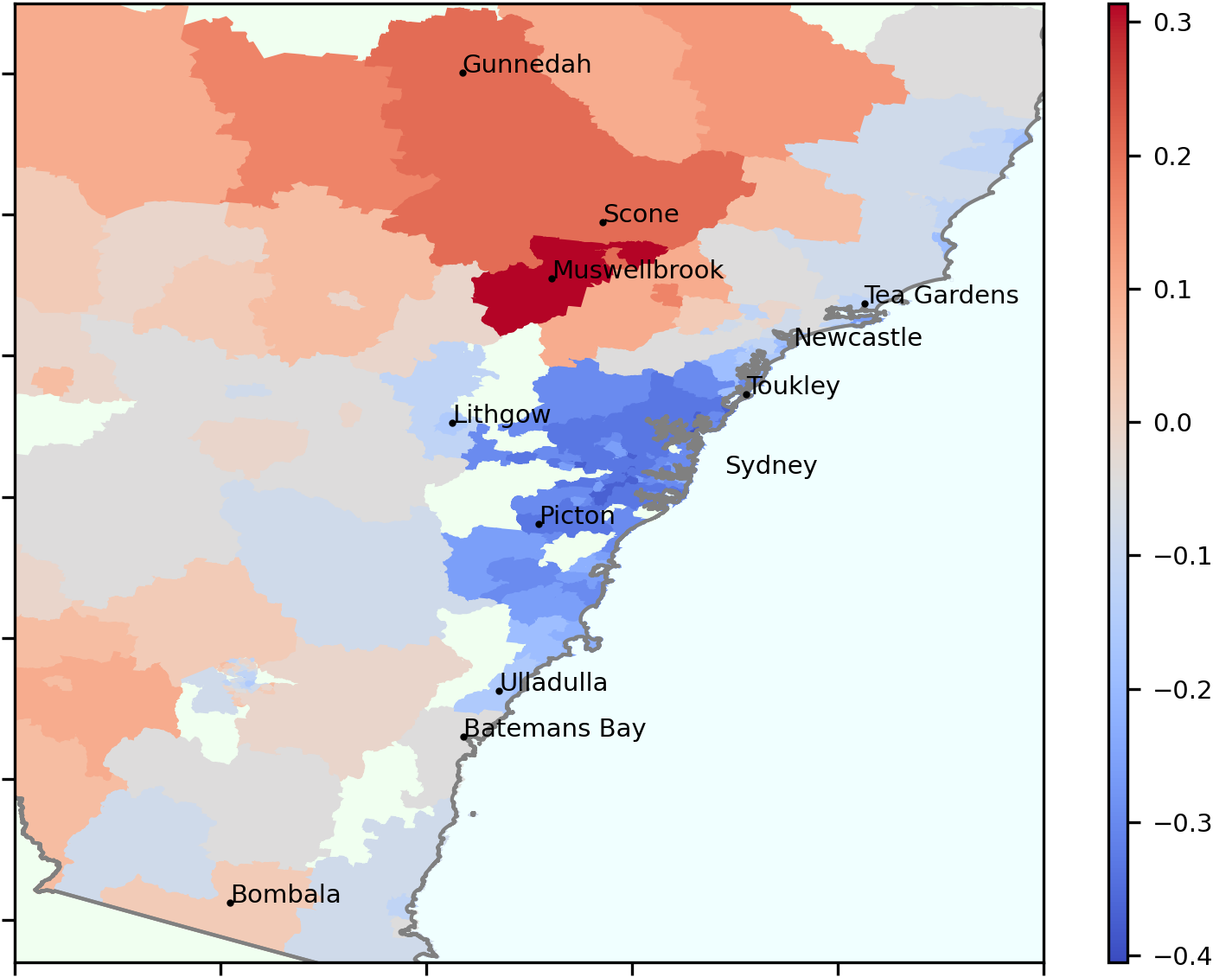}
\caption{PC 2 scores, associated with mining. The color of each SA2 region reflects its PC 2 score, reflecting mining. Of particular interest is Muswellbrook, an area with mining. Sydney has strong negative values. Weaker negative values extend to Ulladulla in the south and beyond Newcastle to the north.  }
\label{map_coastal_nsw_mining}
\end{wrapfigure}

A closer look (figure~\ref{map_coastal_nsw_mining}) reveals the negative PC 2 scores for Sydney and its surroundings more clearly, revealing that the most strongly negative scores only occur in Greater Sydney and its immediate surrounds.
The significance of this remains unclear at this stage. Interpreting strongly negative PC 2 scores as reflecting larger cities appears to conflict with the near-zero score of Brisbane, the third-largest city in Australia. This may result from the mixture of economic activity there, having characteristics in common with Sydney and Melbourne, but also experiencing an influence of the wider QLD mining economy, perhaps leading to cancelling effects. As these considerations remain speculative at this stage, this report will not interpret the negative PC 2 scores, and further study is needed to determine the underlying economic drivers of this effect.

\begin{wrapfigure}{r}{0.7\textwidth} 
    \centering
\includegraphics[width=0.6\textwidth]{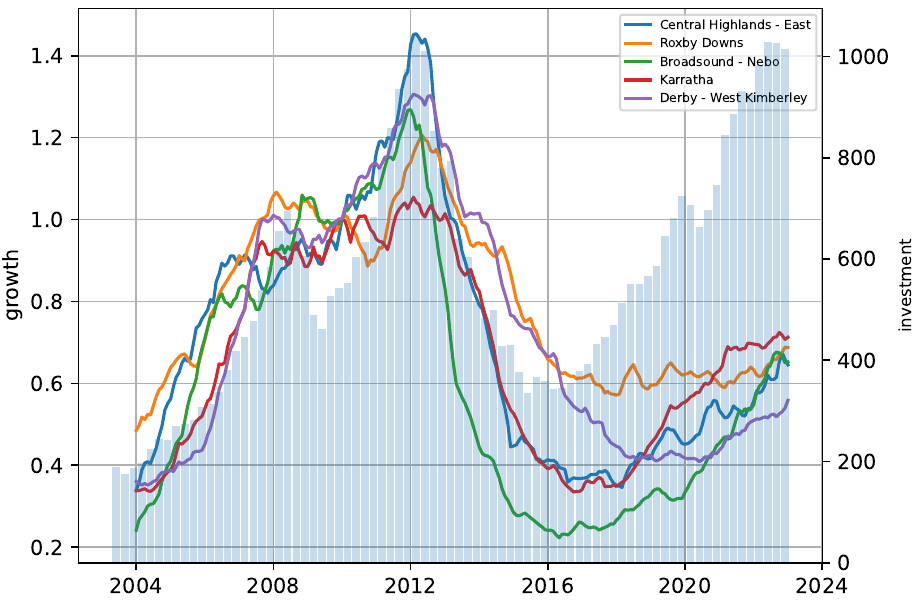}
\caption{Price indexes in real terms for several regions with a high score for PC 2 (mining). The indexes are initialized at 1 in January 2010. 
    The right-hand-side $y$-axis shows mining Expenditure in millions of dollars (Source: ABS 2023, 8412.0 Mineral and Petroleum Exploration, Australia). Real house prices peaked in 2012 at the same time as the mining investment, and saw a steep decline thereafter. The mining-boom induced price surge coincided with relative stagnation in the other markets. This is an example of extreme volatility.  }
\label{mining_series_report}
\end{wrapfigure}

As an example of the ``resources component'', Figure~\ref{mining_series_report} shows house price indexes for several SA2 regions with a high positive score for the second PC. 

Despite some of the regions being far apart geographically, their behavior in time is similar. Strong growth from 2004 to 2008 is associated with a similarly strong mining investment boom peaking in 2008 and again in 2012-2013 just before the steep decline 
 in mining expenditure and also house prices, marking the end of the mineral resources boom in 2013.  The price evolution is clearly influenced by the pattern of mining investment, see right-hand-side $y$-axis of Figure~\ref{mining_series_report} and PC 2 in Figure~\ref{pcs}. Again, this shows the significance of PC 2 scores as a housing market indicator for mining.

 The PC 2 scores calculated from the PCA allows the identification of these regions in an objective and quantitative way based on their price index behavior alone. 
 This volatility underscores the risks associated with resource-dependent markets, where housing affordability and price cycles are directly linked to commodity-driven employment fluctuations. \cite{yanotti2023} also highlight that these regions experience prolonged price stagnations after boom phases, reinforcing the need for caution when considering housing investments in these markets.

 \begin{wrapfigure}{l}{0.7\textwidth} 
    \centering
\includegraphics[width=0.6\textwidth]{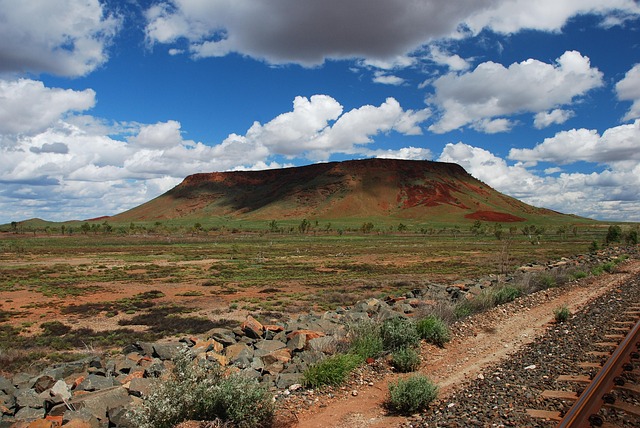}
\caption{Near Karratha.  }
\label{illustration1}
\end{wrapfigure}

\clearpage
\section{Mandurah: a lifestyle region influenced by mining}

So far, PC scores have been described individually. However, some regions score high on both PC 2 and PC 3 (see table \ref{table:eof3}). 
There are interesting regions in WA with high PC~3 and high PC~2 scores, indicating both mining and lifestyle characteristics. 
Mandurah, while reachable from Perth by train, is outside the larger urban environment and an example of such dual-characteristics regions.

 \begin{wrapfigure}{l}{0.7\textwidth} 
    \centering
\includegraphics[width=0.6\textwidth]{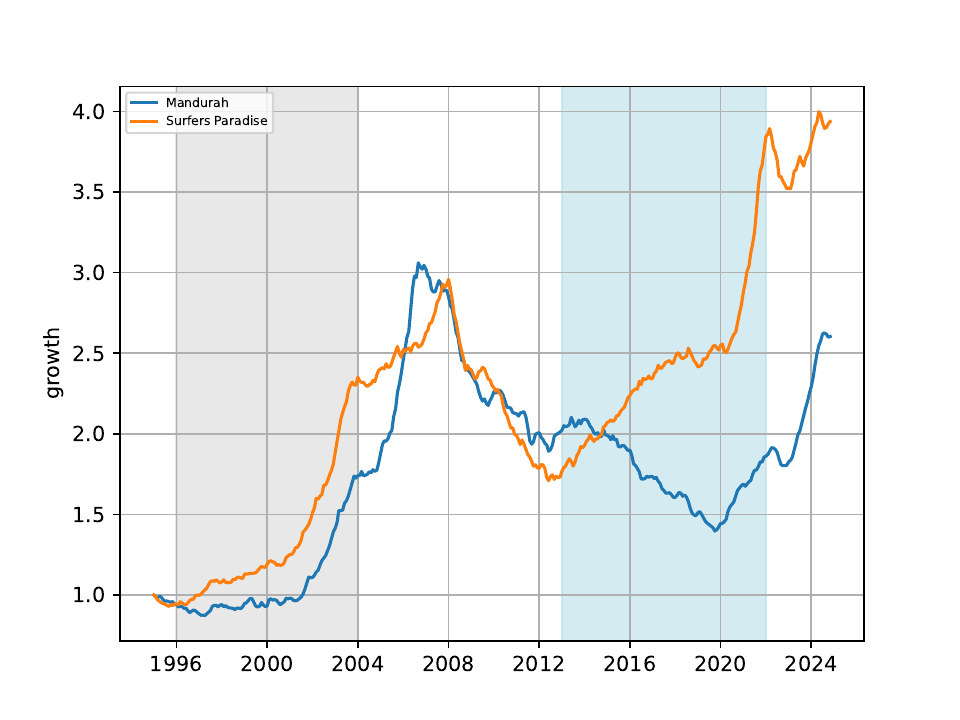}
\caption{CPI-corrected price index for Mandurah and Surfers Paradise. Mandurah, WA, scores high on PC 2, mining, and PC 3, lifestyle, whereas Surfers Paradise only scores high on PC 3, lifestyle, and low on PC 2. The mining influence is expressed in the more prolonged decline after the first boom, ending in 2006. The significant post-2020 price surge and the absence of a 2012 peak are consistent with the lifestyle character of Mandurah.  }
\label{fig_hybrid}
\end{wrapfigure}

There are various newspaper articles that support a mining-lifestyle score for Mandurah, such as one in the Mandurah Mail in 2014 that is titled: ``From mining pull to lifestyle push: Mandurah ranks top.'' Located in the mining-heavy Peel region, Mandurah experiences a significant influence from the resource industry. To complement this, the scenic beauty of the Mandurah Estuary and Peel Inlet, the waterways and proximity to the ocean are all factors that provide Mandurah with a lifestyle appeal, as confirmed by the PC 3 score.
 
To examine the effect of mining within a lifestyle setting, figure~\ref{fig_hybrid} compares the CPI-corrected price index for Mandurah to Surfers Paradise, a lifestyle region that only scores high on PC 3, and low on PC 2. The price surge before the 2006 peak begins later in Mandurah, perhaps paced in part by mining investment. The subsequent decline is more prolonged than in Surfers Paradise: this price fall aspect is shared with other mining regions.  However, unlike the mining regions of figure~\ref{mining_series_report}, there is a post-2020 recovery trend that is more characteristic of the lifestyle regions. Despite this, due to the more prolonged and therefore deeper post-2006 decline, Mandurah price levels have not recovered from their 2006 peak in real terms. However, unlike Mandurah, other mining regions tend to peak at the end of the mining boom in 2012, with a more rapid subsequent decline. These considerations illustrate the possibility of describing local markets in terms of multiple PC score aspects.

\chapter{Concluding remarks}

This report demonstrates how Principal Component Analysis (PCA) provides a transparent, data-driven framework for classifying housing markets based on price trends. It offers objective house price data-driven market classifications and insights into local dynamics. The mining boom left a lasting imprint on resource-rich regions, driving sharp price increases followed by prolonged declines, underscoring the risks associated with resource-dependent markets.
Lifestyle regions, driven by affordability, holiday appeal and migration from urban centers, exhibit cyclical growth patterns. These markets lag behind cities like Sydney during booms but catch up rapidly after a price gap has developed, leading to heightened volatility and long periods of stagnation. This dynamic poses risks for homeowners and investors while creating opportunities for those who understand these patterns.

Many of the regions identified by high lifestyle scores are expected to also score high on external measures of lifestyle desirability, such as the proportion of holiday homes, proximity to the beach and popularity as a holiday destination. 
Similarly, the PCA calculation of the mining score is not based on any model inputs relating to the resource industry, yet it provides a good indicator of the influence of mining in a region. The dominance of these first few trends suggests that these phenomena have had a pronounced impact on the Australian housing market when viewed in geographical terms.

This motivates future work that examines whether PCA scores could serve a purpose as direct indicators of some of these socioeconomic factors, beyond the analysis of solely real estate. For instance, the PC 2 score could serve as a proxy for certain mining activity. Also, future study of the effect of mining on the eastern states urban areas via the broader economy could also shed light on the negative PC 2 scores there. A limitation of this study is that it identifies only two useful scores, as the higher and less dominant PC trends (4, 5, etc.) are more difficult to link to specific socio-economic phenomena at this stage. Future studies could attempt to identify what other types of PCA scores could be constructed.

Although some of the housing market dynamics described in this report are well known, its contribution lies in providing new quantitative tools to study this, allowing the identification of markets that tend to move in similar directions over prolonged periods of time.
The objective characterization of markets through PC scores supports a nuanced understanding of risks and opportunities, enabling more informed investment decisions.  For instance, a sharp annual increase in city prices may anticipate a sharper increase in the lifestyle markets, whereas the investor may need to anticipate a prolonged stagnation at the end of a boom in these regions. For mining region housing markets, this report confirms their cyclical nature, strongly linked to the resources sector.

For real estate businesses, these insights present opportunities to identify emerging markets and refine targeting strategies.  For investors, the findings emphasize the importance of diversification, especially in managing the interplay between urban equity flows and regional market growth. 

Additionally, PCA-derived scores allow real estate businesses to tailor advertising to regional dynamics, enhancing campaign effectiveness. This is borne out by the word frequency counts in this report, that show that local real estate agents increasingly emphasize the lifestyle and holiday appeal when selling a home in a lifestyle region. Based on house price growth patterns alone, the PC scores provide a quantitative guide to determine where and when such advertising may be most effective.

\clearpage
\begingroup
\setlength\bibsep{0.5pt}

\endgroup


\appendix

\chapter{Methodology in More Detail}
\label{method_detail}

\section{Local price indexes}
The methodology takes the seminal work of Bailey, Muth and Nourse \cite{bailey:1963} on the repeat sales methodology as a starting point, and involves the application of PCA to the collection of all local sparse repeat sales house price indexes for the SA2 regions in Australia, $p$ in total. 

In this section, this focus is on the local price indexes that the PCA is later applied to.
The indexes are calculated at once for an entire SA4 region. To put this in perspective, each such region contains on average 25 SA2 regions.
The indexes fit on SA2 regions by themselves would be volatile. However, fitting them together at once and using spatio-temporal regularization mitigates this significantly. 

In more detail, the sparse repeat sales index regression is given by:

\begin{equation}
\boldsymbol{y} = D^{(st)} \boldsymbol{\alpha} + D^{(t)} \boldsymbol{\mu} + \boldsymbol{\varepsilon},
\label{eq:oneregion}
\end{equation}

for a spatio-temporal selection matrix $D^{(st)}$ that selects the region and the time. The vector $\boldsymbol{\alpha}$ contains all local trends in sequence: this can be written as $\boldsymbol{\alpha} = \text{vec}(H)$ for a $T \times p$ matrix $H$ that contains the local trends as columns, written as $H_{.r}$, indicating column $r$. The vec operation concatenates these local trends in succession to allow the application of the spatio-temporal selection matrix.  A main trend $\boldsymbol{\mu}$ is computed so that the local price index for region $r$ is $\boldsymbol{\mu} + H_{.r}$. 
Each element of the vector $\boldsymbol{y}$ contains the difference in log price between subsequent sales of the same house. The temporal design matrix $D^{(t)}$ has column entries taking the value $-1$ at the first time of purchase, $1$ at the time of the repeat sale and $0$ otherwise. This pattern also applies to time in $D^{(st)}$, where the second sale encodes a $1$ and the first a $-1$: it is given by $D^{(st)} = D^{(s)}*D^{(t)}$, the row-wise Kronecker product, for a spatial selection matrix $D^{(s)}$ indicating the region. Finally, $\epsilon_i$ an idiosyncratic error.

Using spatio-temporal regularization, local index values at subsequent times or neighboring regions are able to influence each other. This is made possible by assuming a Gaussian distribution for differences in index values across time and location. To model the relationships between locations and times, a network, or graph, structure is established. The simple networks shown in Figure~\ref{fig:cartoon} are an example, where edges are shown as the lines between the nodes, shown in circles or squares. 
The nodes could represent regions, as in the green graph, and edges could indicate that regions share a border. 
Similarly, time is represented by a linear graph where subsequent times are connected by edges, such as the blue graph on the left in Fig~\ref{fig:cartoon}. These undirected graphs are summarized by a special type of matrix $L$ called the graph Laplacian. It is given by $L=D-W$, with $D$ the degree matrix, a diagonal matrix containing the degree of each graph node, and $W$ the connectivity matrix, containing a $1$ for each pair of connected nodes and $0$ otherwise. So for the linear graph shown (blue), the diagonal matrix $D$ has $(1,2,2,1)$ on the diagonal, and $W$ a $1$ for row $i$ and column $i+1$ for $i=1,2,3$ and vice versa, and $0$ elsewhere.

 The regularization amounts to a transformed or generalized ridge regression, placing a penalty on the \emph{difference} of the price index at adjacent times and also for neighboring regions. As a ridge regression, it naturally handles time periods and regions for which data are missing by imputing interpolated values, and reduces idiosyncratic noise. The regression penalty is of the form:

\begin{subequations}
    \begin{align}
  \|\boldsymbol{y} - D^{(st)} \boldsymbol{\alpha} - D^{(t)} \boldsymbol{\mu} \|_2^2 +\gamma_{st}\boldsymbol{\alpha}'{L^{(st)}}\boldsymbol{\alpha}  + \gamma_t\boldsymbol{\mu}'{L^{(t)}}\boldsymbol{\mu} 
    \end{align}
\label{eq:penalty_general}
\end{subequations}

 where $\gamma^{(st)}$ controls the extent of regularization for the regional trends, and $\gamma^{(t)}$ for the main trend. The graph Laplacian matrices are $L^{(st)}$ for location and time, and $L^{(t)}$ for time alone. Here, the spatio-temporal Laplacian is given by the Kronecker product of the spatial and temporal Laplacians: $L^{(st)} = L^{(s)} \otimes L^{(t)}$, where the spatial graph encodes neighboring regions.
The effect of the $\mathbf{L}$ terms is to penalize differences between values of the index in adjacent time periods and regions, encapsulating the prior that differences in trends for neighboring regions should have a centered Gaussian distribution that puts limits on their divergence, where $\text{var}(\alpha)=\text{var}(\varepsilon)/{\gamma^{(st)}}$ and $\text{var}(\mu)=\text{var}(\varepsilon)/{\gamma^{(t)}}$. This reflects the economic intuition that house price changes cannot be
arbitrarily large over short periods (in fact it is a random walk) and adjacent regions, and aligns with observed market behaviors of location substitutability \cite[][]{alonso1964, mills1967}: when prices change in one area, they are likely to change at a similar rate in nearby areas. This model can be estimated by a transformed OLS-based ridge regression in standard software packages like Statsmodels, where an out of sample grid search is applied to find the ridge parameters. Alternatively, the model can be estimated in the open-source probabilistic programming language and Bayesian inference engine Stan, so that the variances can be estimated as part of the probabilistic framework.

\section{PCA}

\begin{figure}[ht]
\begin{center}
\includegraphics[width=0.9\textwidth]{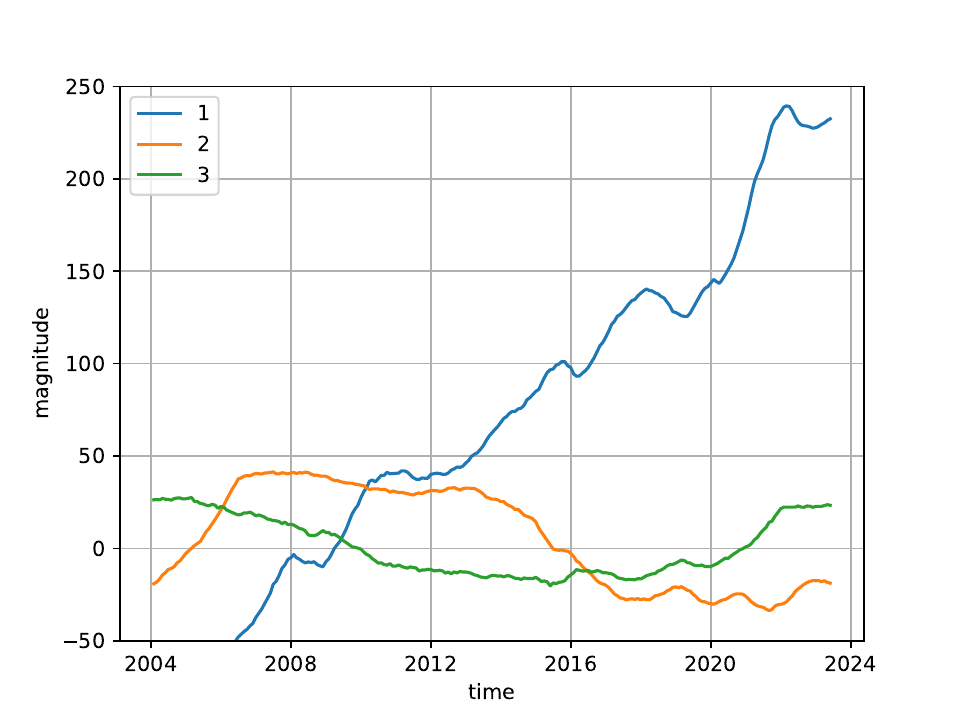}
\end{center}
\vspace{0cm}
\caption{The first three principal components for all SA2 regions in Australia.}
\begin{tablenotes}
      \footnotesize
	   \item \textit{Notes}: The vertical scale has been truncated to bring out components 2 and 3. The first PC captures the main trend, the second one is associated with the mining boom, and the third one with a lifestyle phenomenon. The local indexes are correlated to various degrees with these main trends and, in turn, can be expressed as their weighted sum. These weights are the PC scores.   
\end{tablenotes}
\label{pcs}
\end{figure}

The second step is principal component analysis (PCA) applied to the set of regional price indexes  obtained from the first step. This identifies the main uncorrelated underlying trends (time series) in the Australian market, ranked in descending order of importance (magnitude), as shown in figure~\ref{pcs}. In the Australian context, the most important trend (blue, labelled ``1'') represents the main national trend, the second trend will turn out to be related to the mining boom and the third to the lifestyle property market.

PCA also yields a score for each region, showing how strongly it is influenced by each main trend. This is because each regional trend can be calculated as a weighted sum of these main trends, where the weights are used as scores. This is solely based on the price index behavior of the local housing market.
PCA is also a way of smoothing data, in this case based on similarities between potentially distant markets. This is achieved by truncating the weighted sums of the trends \cite{jolliffe1986}.

This is described in detail by \cite{jolliffe1986}, and in technical terms, if the $T \times p$ matrix $\boldsymbol{F}=[\boldsymbol{\mu}_1, \ldots, \boldsymbol{\mu}_p]$ holds one of the $p$ regional price indexes in each column, each a time series of $T$ points in time, PCA is used on the dataset $\boldsymbol{F}$ to summarize these regional price indexes in terms of only a few (linearly transformed) time series $\boldsymbol{z}_k$ that capture most of the variance and so retain most of the information. This is done by computing the covariance matrix $\boldsymbol{R}=\boldsymbol{F}'\boldsymbol{F}$ and its associated eigenvectors. The weighted sum referred to earlier expresses the local price indexes in terms of the newly constructed  $\boldsymbol{z}_k$ as:

\begin{equation}
    \boldsymbol{\mu}_r = \sum_{k=1}^{p}\alpha_{kr} \boldsymbol{z}_{k}.
\label{eq:mu_transform}
\end{equation}

More specifically, the column vectors $\boldsymbol{\alpha}_k$ are the eigenvectors belonging to the eigenvalues $\lambda_k$ ranked in descending order.
The weights matrix $\alpha_{kr}$ contains the scores that indicate how strongly each local market is influenced by each local trend. The $\boldsymbol{z}_k$ time series are ranking according to descending importance: they are the Principal Components (PCs) and usually the first few of them represent main trends in the granular index data. Only the first few of these PCs are needed in practical analysis. Smoothing is achieved by truncating this sum, as most of the time series can be reconstructed from them and the higher trends represent noise. As the spatio-temporal regularization of the first step is already very effective, the second PCA step may not be needed to achieve smooth indexes. Research on the mitigating effect of the second (PCA) step on revision volatility is ongoing.

In summary, the PCA serves to provide scores to the regions indicating how strongly they couple to each main trend. Also, it provides a way of obtaining smoother local price indexes by truncation.

\section{A model pipeline}

\begin{figure}[H]
    \begin{center}
    \includegraphics[width=.8\textwidth]{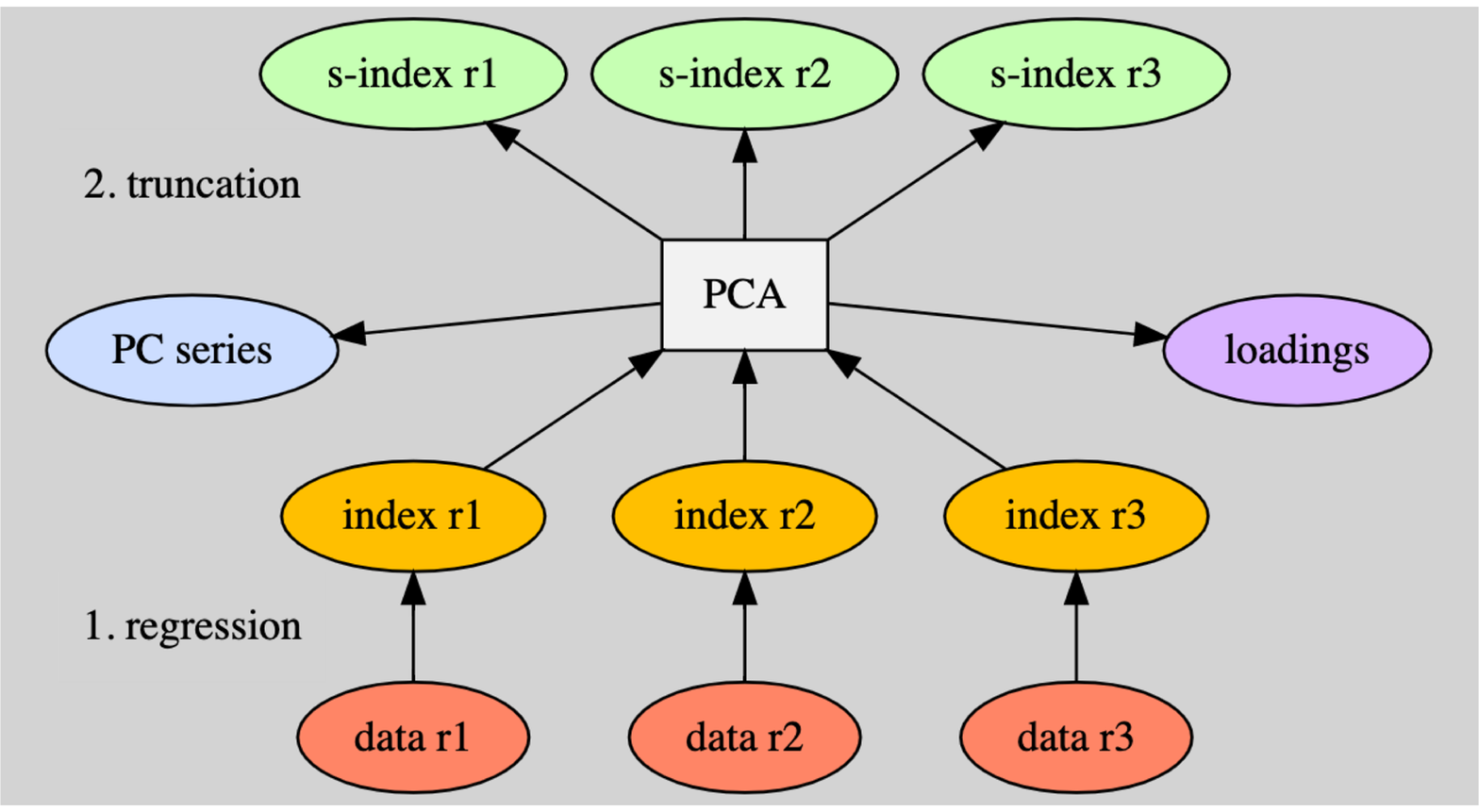}
    \end{center}
    \caption{Schematic of the entire PCA index model pipeline for three regions $r_1, r_2, r_3$.}
    \label{diagram}
    \begin{tablenotes}
        \footnotesize
        \item \textit{Notes}: In step 1, sales data (red), is used to perform regularized regression, yielding a collection of still relatively noisy indexes (orange) to which PCA is applied collectively (gray box) in step 2. Truncation yields the smoothed truncated index time series indicated by ``s-index'' (green). PCA also yields the PC series (main trends, blue) and the loadings (geographical patterns, purple), useful for analysis. 
    \end{tablenotes}
\end{figure}

Smooth granular indexes can be produced in a two-step procedure, as indicated schematically in Figure~\ref{diagram}:
\begin{enumerate} 
    \item Estimate regional price indexes.
    \item Apply PCA on the estimated regional price indexes, and obtain PCA de-noised regional price indexes.
\end{enumerate}

Step 1 provides estimated regional price indexes for the $T$ time steps and $p$ regions, the orange boxes of Figure~\ref{diagram}.

Step 2 applies PCA to the collection of regional indexes.
PCA requires that there be no missing values, and indeed this is guaranteed by the spatio-temporal regularization used for the repeat sales indexes in this report, yielding balanced panel index data. 
The ``finished product'' of the model pipeline in Figure~\ref{diagram} expresses local indexes in terms of a few main principal components. This is to yield smooth indexes, and is not necessary for the analysis here, as this report relates to the PC scores for the regions. If sufficiently smooth and accurate local indexes can be produced, the PCA step may not be needed for smoothing: this depends on the use-case and data.

\section{Large Tables}

Table~\ref{table:eof3} shows the top nine SA2 regions by PC 3 (lifestyle) scores for each greater capital city statistical area. 

\begin{table}[H]
	\centering
\scalebox{0.7}{
\begin{tabular}{llrrr}
\toprule
 Major city &                           SA2 name &  PC 1 &  PC 2 &  PC 3 \\
\midrule
Rest of NSW &           Tea Gardens - Hawks Nest &              0.87 &     -0.05 &      0.42 \\
Rest of NSW &            Forster-Tuncurry Region &              0.91 &     -0.06 &      0.34 \\
Rest of NSW &               Nelson Bay Peninsula &              0.91 &     -0.19 &      0.32 \\
Rest of NSW &                            Forster &              0.92 &     -0.14 &      0.32 \\
Rest of NSW &                           Tuncurry &              0.90 &     -0.17 &      0.30 \\
Rest of NSW &               Batemans Bay - South &              0.94 &     -0.02 &      0.29 \\
Rest of NSW &                           Anna Bay &              0.94 &     -0.07 &      0.29 \\
Rest of NSW &                   Ulladulla Region &              0.94 &     -0.13 &      0.28 \\
Rest of NSW & Old Bar - Manning Point - Red Head &              0.93 &     -0.06 &      0.28 \\
Rest of TAS &                        George Town &              0.93 &      0.20 &      0.21 \\
Rest of TAS &              St Helens - Scamander &              0.95 &      0.13 &      0.19 \\
Rest of QLD &                       Port Douglas &              0.83 &      0.00 &      0.34 \\
Rest of QLD &                           Daintree &              0.77 &      0.29 &      0.33 \\
Rest of QLD &      Torquay - Scarness - Kawungan &              0.92 &      0.23 &      0.30 \\
Rest of QLD &                   Surfers Paradise &              0.93 &      0.02 &      0.29 \\
Rest of QLD &                           Cooloola &              0.89 &      0.26 &      0.29 \\
Rest of QLD &               Booral - River Heads &              0.91 &      0.27 &      0.29 \\
Rest of QLD &                 Urangan - Wondunna &              0.92 &      0.23 &      0.29 \\
Rest of QLD &        Craignish - Dundowran Beach &              0.92 &      0.25 &      0.28 \\
Rest of QLD &            Caloundra - Kings Beach &              0.95 &      0.04 &      0.28 \\
 Rest of SA &                          Millicent &              0.90 &      0.29 &      0.25 \\
 Rest of SA &                            Tatiara &              0.90 &      0.27 &      0.23 \\
 Rest of SA &                    Kangaroo Island &              0.94 &      0.21 &      0.19 \\
 Rest of SA &                              Berri &              0.93 &      0.22 &      0.19 \\
 Rest of SA &                    Kingston - Robe &              0.94 &      0.20 &      0.18 \\
 Rest of SA &                      Victor Harbor &              0.96 &      0.14 &      0.18 \\
   Brisbane &                    Redland Islands &              0.89 &      0.27 &      0.33 \\
   Brisbane &                      Logan Village &              0.53 &      0.20 &      0.29 \\
   Brisbane &                      Bribie Island &              0.95 &      0.09 &      0.26 \\
   Brisbane &                          Jimboomba &              0.80 &      0.30 &      0.24 \\
   Brisbane &        Beachmere - Sandstone Point &              0.95 &      0.17 &      0.22 \\
   Brisbane &                      Deception Bay &              0.96 &      0.14 &      0.20 \\
   Brisbane &                             Kilcoy &              0.91 &      0.26 &      0.19 \\
   Brisbane &                        Redland Bay &              0.96 &      0.13 &      0.19 \\
   Brisbane &                         Caboolture &              0.95 &      0.20 &      0.18 \\
Rest of VIC &                    Hamilton (Vic.) &              0.94 &      0.17 &      0.21 \\
Rest of VIC &                           Portland &              0.95 &      0.18 &      0.18 \\
     Sydney & Budgewoi - Buff Point - Halekulani &              0.94 &     -0.26 &      0.20 \\
     Sydney &               Warnervale - Wadalba &              0.96 &     -0.17 &      0.19 \\
     Sydney &     Lake Munmorah - Mannering Park &              0.95 &     -0.25 &      0.19 \\
     Sydney &                       The Entrance &              0.91 &     -0.35 &      0.18 \\
     Sydney &              Blue Haven - San Remo &              0.94 &     -0.25 &      0.18 \\
     Sydney &       Summerland Point - Gwandalan &              0.95 &     -0.24 &      0.18 \\
      Perth &               Dawesville - Bouvard &              0.85 &      0.42 &      0.26 \\
      Perth &                  Falcon - Wannanup &              0.85 &      0.42 &      0.25 \\
      Perth &                           Mandurah &              0.84 &      0.47 &      0.20 \\
      Perth &                   Mandurah - North &              0.86 &      0.44 &      0.19 \\
      Perth &                   Mandurah - South &              0.86 &      0.45 &      0.18 \\
 Rest of WA &                   Busselton Region &              0.90 &      0.31 &      0.19 \\
\bottomrule
\end{tabular}
}
	\caption{Scores for principal component 3, associated with lifestyle.}
	\label{table:eof3}
     \begin{tablenotes}
     \footnotesize
       \item \textit{Notes}:    
       Top nine SA2 regions by PC 3 (lifestyle) scores for each greater capital city statistical area.  
       Only regions with more than 1,000 sales in the index period and with a PC 3 score of more than 0.18 are included. 
       This threshold is chosen such that Sydney is included. These rankings are not according to price performance or external economic data. Instead, they are solely based on the PC scores derived from general index characteristics.
    \end{tablenotes}

\end{table}

\end{document}